\documentclass[preprint,12pt]{aastex}
\def\simlt{\lower.5ex\hbox{$\; \buildrel < \over \sim \;$}}
\def\simgt{\lower.5ex\hbox{$\; \buildrel > \over \sim \;$}}

\def\cm3{{\rm cm^{-3}}}
\def\kms{km s$^{-1}$}
\def\msol{{$M_\odot$}}
\def\lsol{{$L_\odot$}}

\def\alphaJ{{\alpha_{2000}}}
\def\deltaJ{{\delta_{2000}}}

\def\g54{G54.1+0.3}
\def\akari{{\em AKARI}\ }
\def\spitzer{{\em Spitzer}}
\def\iras{{\em IRAS}}
\def\msx{{\em MSX}}

\def\um{$\mu$m}

\def\fe164{[Fe II] 1.644~$\mu$m}

\def\h212{H$_2$ 2.122~$\mu$m}

\def\tmass{{\em 2MASS}}

\def\kms{km s$^{-1}$}
\def\msol{{$M_\odot$}}
\def\lsol{{$L_\odot$}}

\def\alphaJ{{\alpha_{2000}}}
\def\deltaJ{{\delta_{2000}}}
\def\simlt{\lower.5ex\hbox{$\; \buildrel < \over \sim \;$}}
\def\simgt{\lower.5ex\hbox{$\; \buildrel > \over \sim \;$}}
\def\spitzer{{\em Spitzer}}
\def\akari{{\em AKARI}\ }
\def\aste{{\em ASTE}\ }
\def\irasx{IRAS 15099$-$5856}
\def\msh{MSH 15$-$5{\em 2}}
\def\um{$\mu$m}
\def\lccs{L_{\rm IRS1}}

\def\dfour{d_{\rm 4}}

\def\neratio{[Ne III]/[Ne II]}

\newcommand{\beq}	{\begin{equation}}
\newcommand{\eeq}	{\end{equation}}
\newcommand{\beqa}{\begin{eqnarray}}
\newcommand{\eeqa}{\end{eqnarray}}
\newcommand{\beqs}	{\begin{displaymath}}
\newcommand{\eeqs}	{\end{displaymath}}
\newcommand{\beqas}	{\begin{eqnarray*}}
\newcommand{\eeqas}	{\end{eqnarray*}}

\def\bit{\begin{itemize}}
\def\eit{\end{itemize}}

\def\ga{\simgt}

\font\tenbi=cmmib10 
\newfam\bifam  \textfont\bifam=\tenbi

\font\tenbr=cmbx10
\newfam\brfam  \textfont\brfam=\tenbr

\shortauthors{Koo et al.}
\shorttitle{IRAS 15099$-$5856}

\begin{document}
\title{IRAS 15099$-$5856: Remarkable Mid-Infrared Source with Prominent Crystalline Silicate
Emission Embedded in the Supernova Remnant MSH15$-$5{\em 2}}

\author{Bon-Chul Koo\altaffilmark{1}, 
Christopher F. McKee\altaffilmark{2},
Kyung-Won Suh\altaffilmark{3},
Dae-Sik Moon\altaffilmark{4}, 
Takashi Onaka\altaffilmark{5}, 
Michael, G. Burton\altaffilmark{6}, 
Masaaki Hiramatsu\altaffilmark{7}, 
Michael S. Bessell\altaffilmark{8},
B. M. Gaensler\altaffilmark{9},
Hyun-Jeong Kim\altaffilmark{1},
Jae-Joon Lee\altaffilmark{10, 11}, 
Woong-Seob Jeong\altaffilmark{11}, 
Ho-Gyu Lee\altaffilmark{4}, 
Myungshin Im\altaffilmark{1},
Ken'ichi Tatematsu\altaffilmark{12},
Kotaro Kohno\altaffilmark{13},
Ryohei Kawabe\altaffilmark{12}, Hajime Ezawa\altaffilmark{12},
Grant Wilson\altaffilmark{14},  Min S. Yun\altaffilmark{14}
David H. Hughes\altaffilmark{15}
}

\altaffiltext{1}{Department of Physics and Astronomy, Seoul National University, Seoul 151-742, Korea;
koo@astrohi.snu.ac.kr}
\altaffiltext{2}{Departments of Physics and Astronomy, University of California, Berkeley, CA94720, USA; and Laboratoire d'Etudes du Rayonnement et de la Mati\`ere en Astrophysique, LERMA-LRA, Ecole Normale Superieure, 24 rue Lhomond, 75005 Paris, 
France}
\altaffiltext{3}{Department of Astronomy and Space Science, Chungbuk National University, Cheingju-City 361-763, Korea}
\altaffiltext{4}{Department of Astronomy \& Astrophysics, University of Toronto, Toronto ON M5S 3H4, Canada}
\altaffiltext{5}{Department of Astronomy,
University of Tokyo, Bunkyo-ku, Tokyo 113-0033, Japan}
\altaffiltext{6}{School of Physics, University of New South Wales, Sydney, New South Wales 2052, Australia}
\altaffiltext{7}{Academia Sinica Institute of Astronomy and Astrophysics,
 P.O. Box 23-141, Taipei 10617, Taiwan}
\altaffiltext{8}{Research School of Astronomy and Astrophysics, 
Mount Stromlo Observatory, Austalia}
\altaffiltext{9}{Sydney Institute for Astronomy, 
School of Physics, The University of Sydney, NSW 2006, Australia}
\altaffiltext{10}{Astronomy \& Astrophysics Department, Pennsylvania State University, University Park PA16802, USA}
\altaffiltext{11}{Korea Astronomy and Space Science Institute, 61-1, Whaam-dong, Yuseong-gu, 
Daejeon 305-348, Korea}
\altaffiltext{12}{National Astronomical Observatory of Japan, 2-21-1, Osawa, Mitaka, Tokyo 181-8588, Japan}
\altaffiltext{13}{Institute of Astronomy, The University of Tokyo, 2-21-1 Osawa,
Mitaka, Tokyo 181-0015, Japan}
\altaffiltext{14}{Department of Astronomy, University of Massachusetts, Amherst,
MA 01003, USA}
\altaffiltext{15}{Instituto Nacional de Astrofisica, Optica y Electronica,
Tonantzintla, Aptdo. Postal 51 y 216, 72000 Puebla, Pue., Mexico}

\begin{abstract}

We report new mid-infrared observations of the remarkable object \irasx\ 
using the space telescopes \akari\ and \spitzer, which demonstrate the presence of 
prominent crystalline silicate emission in this bright source.
\irasx\ has a complex
morphology with a bright central compact source (IRS1)
surrounded by knots, spurs, and
several extended ($\sim 4'$) arc-like filaments. 
The source is seen only at $\ge 10$~\um. 
The \spitzer\ 
mid-infrared (MIR) spectrum of IRS1 shows 
prominent emission features from Mg-rich crystalline silicates,
strong [Ne II] 12.81 \um\ and several other faint ionic lines.
We model the MIR spectrum as thermal emission from 
dust and compare with the Herbig Be star HD 100546 and 
the luminous blue variable R71, 
which show very similar MIR spectra.
Molecular line observations reveal two molecular clouds 
around the source, but no associated dense molecular cores.
We suggest that IRS1 is heated
by UV radiation from the adjacent O star Muzzio 10
and that its crystalline silicates most likely
originated in a mass outflow from the progenitor 
of the supernova remnant (SNR) \msh.
IRS1, which is embedded in the SNR, 
could have been shielded from the SN blast wave if 
the progenitor was in a close binary system with Muzzio 10.
If \msh\ is a remnant of Type Ib/c supernova (SN Ib/c), as has been previously proposed, 
this would confirm 
the binary model for SN Ib/c.
IRS1 and the associated structures may be the relics of massive
star death, as shaped by the supernova explosion,
the pulsar wind and the intense ionizing radiation
of the embedded O star. 

\end{abstract}


\section{Introduction}

Space infrared observations have revealed distinct spectral features 
due to crystalline silicates in diverse objects such as 
Asymptotic Giant Branch (AGB) stars, 
planetary nebulae, Herbig Ae/Be stars, comets 
and ultraluminous infrared galaxies \citep{hen10}.
Since silicate dust grains in the interstellar medium are
essentially amorphous \citep{kemper04}, 
crystalline silicates must form 
in circumstellar disks and/or outflows of evolved stars or 
young stellar objects (YSOs). 
However, neither the formation process of the crystalline dust nor
its relation to the central stellar source is understood.

In this paper, we report the discovery of
prominent crystalline silicate emission in
\irasx, a bright, mid-infrared compact source
previously detected by {\em Infrared Astronomical Satellite} (\iras) and 
{\em Midcourse Space Experiment} (\msx).
The source is located close
to the pulsar B1509$-$58 in the supernova remnant (SNR) \msh\ 
(G320.4$-$1.2), a young
SNR of complex morphology
at a distance of $5.2\pm 1.4$~kpc \citep[][see \S~5]{arendt91, gaensler99}.
\cite{arendt91} concluded that \irasx\ is heated either by hot plasma associated with
the SNR or by the nearby O star Muzzio 10 \citep{muzzio79}, 
which is at a distance $\sim 4$~kpc \citep{bessell11}. 
We show that the latter explanation is most likely correct and
that the source is probably associated with 
the progenitor of the SNR.

\section{\akari\ Images and Spectral Energy Distribution}

We observed \irasx\ using the
Infrared Camera (IRC) aboard \akari\ on February 21-22, 2007 (ID: 1400761.1 and 1400762.1).
The IRC is equipped with three waveband channels covering the 2.6--26 $\mu$m
wavelength range
with a 10$'$ $\times$ 10$'$ field-of-view (FOV) \citep{onaka07}. 
We used all three channels in IRC02 mode, which
gave six band images, N3, N4, S7, S11, L15, and L24, with
central wavelengths of
3.2, 4.1, 7, 11, 15 and 24 $\mu$m, respectively,
and angular resolutions ranging from $4.''2$ (N3) to $6.''8$ (L24).
The data were processed using the
standard IRC Imaging Data Reduction Pipeline version 070104.
The positional uncertainty ($1\sigma$) is $\le 0.''15$ for N3- and N4-band 
images and $\le 0.''32$ for the longer-wavelength images except the L24-band 
image, which has a relatively large uncertainty 
$(\sim 1'')$ because the astrometric solution 
could not be obtained by matching with
the coordinates of 2MASS stars in the Pipeline. 

The \akari\ observations reveal that 
\irasx\ is a large, MIR source with spectacular morphology 
(Fig. 1). It is composed of 
a bright central compact source (hereafter IRS1), 
a surrounding halo of $\sim 1'$ radius with knots and spurs embedded in it,
and several extended ($\sim 4'$), knotty arc-like filaments. 
The extent of the whole structure is about $10'$ and its total flux is 
$\sim 12$ Jy at 15 \um, 40\% of which is from IRS1. 
The 15 \um\ image in the right frame shows that 
IRS1 has a brightest central part extended along the east-west direction 
and some diffuse emission around it.
There is no \tmass\ counterpart to IRS1.
The source is not seen in the IRC images at wavelengths $\lambda<10$ \um.

Figure~2 shows the one-dimensional
24, 15, and 11~\um\ intensity profiles of IRS1 along the east-west (P.A.=$110^\circ$)
and north-south directions, which are
the two orthogonal directions of the \akari\ IRC.
We have subtracted a
constant background brightness from each profile and
normalized it by its maximum brightness:
background brightness of about 150, 60, and 30~MJy sr$^{-1}$ and
maximum brightness of 3700, 1110, and 30 MJy sr$^{-1}$ for L24,
L15, and S11 profiles, respectively.
For comparison, we also show the point spread function (PSF) of
the \akari\ IRC.
The 15 and 24 \um\ profiles clearly show that IRS1 is
composed of a compact bright central
part (hereafter, the core) and an extended envelope of $\sim 30''$ size.
The envelope appears as a plateau in the north-south profile, which suggests that
it is a separate component from the core.
The core is almost symmetrically elongated along the
east-west direction, while it is barely resolved
in the north-south direction.
The average full-width at half-maximum (FWHM) of the core at 15/24~\um\ 
after deconvolving the PSF is $9.''6\, (0''.1)\times 5.''1\, (1.''1)$
or $0.19 \dfour$~pc~$\times 0.10 \dfour$~pc,
where $\dfour$ is the distance in units of 4 kpc.
(We normalize the distance by 4 kpc, which is the distance
to the Cir OB1 association in this area. See \S~5.)
At 11 \um, the core has a comparable size
($9.''7$) along the east-west direction, but its center is 
slightly ($1.''5$) shifted to the west. Along the north-south direction, 
it appears somewhat more extended ($6.''4$), but this could be due to the 
contribution from nearby faint stellar sources (see Fig. 1).
The local peak at $-13''$ in the 11~\um\ north-south profile
is due to Muzzio 10.

Table 1 summarizes the observed fluxes of \irasx, where we also list
the corresponding ``color-corrected'' fluxes, i.e., 
the fluxes corrected for the spectral slope over the passbands.
The color correction has been made 
by assuming modified black-body emission at 75 K that fits 
the overall shape of the spectral energy distribution (SED). 
But as we show in \S~3, the MIR emission at $\le 35$~\um\ 
is dominated by spectral bumps from crystalline and metallic dusts,
so that the color-corrected fluxes of those bands should be used with caution.   
The AKARI IRC fluxes are derived by summing the fluxes 
within a circle of $25''$ radius 
centered on the brightness peak; the contribution from 
the O star Muzzio 10 is excluded. 
The \iras\ and \msx\ fluxes are from the point source catalogs and they 
roughly agree with the \akari\ IRC fluxes.
At wavelengths shorter than the \akari\ IRC bands, 
we identified several point sources of 17$^{\rm {th}}$--20$^{\rm {th}}$ mag
around IRS1 in the $K_s$ image of the source obtained with
the Magellan 6.5 m telescope \citep{kaplan06}.
They are separated from the 15 \um\ peak position 
by $\ge 1''.1$, which is greater than 
the $3\sigma$ positional uncertainty of the 15~\um\ image.  
We use the $3\sigma$ upper limit (20$^{\rm {th}}$ mag or 0.007 mJy) 
of the Magellan observations
as an upper limit at 2.15 $\mu$m.
At longer wavelengths, 
we identified the source in the \akari\ FIS 
(Far-Infrared Surveyor) 
catalog at 65 and 90 $\mu$m but not at 140 or 160 $\mu$m. 
With a spatial resolution of 37$''$--61$''$, 
the FIS does not resolve IRS1.
We also carried out 1.1 mm continuum observations using the 144-element 
bolometer array AzTEC \citep{wilson08}
on the 10-m telescope {\em Atacama Submillimeter Telescope Experiment} 
\citep[\aste;][]{ezawa04} 
in October 2008. The beam size and field of view were $28''$ and $7.'5$, respectively. 
We obtained 11 maps of a $25' \times 25'$ area in a raster scan
at $90''$ per sec, and coadded the individual maps to obtain the 
final map with a pixel size of $6''$.
The total on-source observing time was 3 hrs and the per-pixel integration time
was 10 s.
The absolute flux calibration was done by observing Uranus.
\irasx\ was not detected with an upper limit of 30 mJy ($2\sigma$).

Figure 3 shows the SED of IRS1. 
The flux of IRS1 increases steeply from short wavelengths 
to $25$~\um, and remains flat at $\sim 30$~Jy to 90~\um. 
The MIR ($11-25$ \um) spectral index is
$\alpha=-7.7$ ($F_\nu\propto \nu^{\alpha}$).
The SED can be approximately fitted by emission from $2.4\times 10^{-3} \dfour^2$~\msol\ of dust at 79 K using Draine's carbonaceous-silicate opacity model for interstellar dust with $R_V=5.5$ \citep{draine03}. The fit, however, is not good, with a large reduced $\chi^2 (=9.4)$. This is because the flux at $\simlt 35$~\um\ is mainly from dust of circumstellar origin with several distinct opacities (\S~3).
In \S~3, we discuss the model based on the \spitzer\ IRS spectra, where
we show that the dust mass is about four times larger because the dust 
temperature is lower.
The total IR luminosity of IRS1 is
$\lccs \approx 2.0\times 10^3 \dfour^2$~\lsol.

\section{\spitzer\ IRS Spectrum and Crystalline Silicate Emission}

We carried out \spitzer\ IRS staring observations of IRS1 of \irasx\ 
with the two low-resolution modules; the Short-Low (SL) module  
(5.2--14.5 \um) and the Long-Low (LL) module (14.0--38.0 \um). 
The observations were done on October 3, 2008 (ID 26318080 and 26318336).
We followed the standard IRS staring mode, where 
the spectra were obtained by placing the 
source at two locations for each slit, located 
$1/3$ and $2/3$ of the way along the slits' length (Fig. 1). 
At each location, 2--3 cycles of different ramp durations (6 s to 60 s) 
were carried out.
We use the coadded 2-D spectra provided by 
the Spitzer Science Center (SSC), which are averaged 
basic calibration data (BCD) FITS files 
processed by the pipeline version 17. 
For the data analysis, we use the Spitzer IRS Custom Extraction (SPICE) software 
v2.2 provided by the SSC. 
The IRS1 is extended, so that the flux that misses the slit, 
i.e. ``slit-loss", needs to be properly accounted for the flux calibration
(see the SSC IRS calibration web page on extended 
sources\footnote{http://ssc.spitzer.caltech.edu/irs/calib/extended\_sources/}). 
We derive the slit-loss correction factor (SLCF) at 11, 15, and 24 $\mu$m by 
calculating the fraction of the flux within the SL and LL slits on 
the \akari\ images using the Point Spread Function in \cite{sloan03a}.  
The SLCFs over the entire SL and LL 
bands are then obtained by interpolating/extrapolating
these data points.

The \spitzer\ IRS spectrum of IRS1 (Fig. 4) 
increases steeply from $\simlt 0.1$ Jy at $\simlt 13$ \um\ to 
20--30 Jy at $\ge 20$ \um, which is consistent with the broadband SED. 
The spectrum shows two broad, strong
peaks at 23 and 34 \um\ and also weak features at 20 and 27 \um.
Such spectral features are characteristic of 
Mg-rich crystalline silicates, e.g., forsterite (Mg$_2$SiO$_4$), and 
have been observed in AGB stars, YSOs, and 
comets \citep{hen10}.
Indeed, the $\ge 15$~\um-part of the spectrum is
very similar to those of YSOs such as HD 100546 or RECX 5 (Fig. 4).
HD 100546 is the Herbig Be star (B9.5) 
that has an exceptionally large 
crystallinity in the {\em Infrared Space Observatory} (ISO) sample 
of Herbig Ae/Be stars \citep{malfait98, meeus01}.
RECX 5 is an M4-type pre-main-sequence star in the 
$\eta$ Chamaeleontis cluster and shows a very similar spectrum 
to HD 100546 \citep{bouwman05}. 
The spectrum is also quite similar to that of 
luminous blue variables (LBVs) such as R71 (HD 269009)  
in the Large Magellanic Cloud (LMC) (Fig. 4). R71 is 
a well-studied LBV with $T_*=15,000$~K and $L=5\times 10^5$~\lsol\ 
\citep{boyer10}. Its ISO and \spitzer\ IRS spectra show a 
broad bump steeply rising from 15 \um\ and flattening 
around 18 \um, which is very similar to that of IRS1 
\citep{voors99, morris08}. 
This bump had been attributed to the metal oxide 
FeO \citep{voors99}.
It also shows the weak bumps 
due to Mg-rich crystalline silicates.
The closely-matching spectral
features suggest that the grain composition 
should be similar in these sources.
A noticeable difference, however, is the absence of 
crystalline/amorphous silicate emission at 10--11 \um\ in \irasx, 
which is distinct in the other sources.
This indicates that \irasx\ is missing the warm (several hundred K)
silicate dust that the other sources have.
Before closing the comparison to other MIR spectra with similar features, 
we note that dust emission 
has been recently detected toward several Oxygen-rich SNRs, 
e.g., Cas A, G292.0+1.8, and the LMC SNR N132D \citep{rho08, ghavamian09, tappe06}. 
The 17--22 \um\ features in the spectra of these objects have
been attributed to either 
silicate dust newly formed in supernova ejecta 
or swept-up polycyclic aromatic hydrocarbons.
They, however, do not show crystalline silicate features.

Several ionic lines are present in \irasx:
strong [Ne II] 12.81 \um\ 
and weak 
[Ne III] 15.56 \um, [S III] 18.71 \um, and 
[O IV] 25.89/[Fe II] 25.99 \um. 
The central velocity of the [Ne II] line is 
$-160\pm 560$~\kms, which indicates that \irasx\ is not at 
a cosmological distance. 
The flux of the [Ne II] 12.81 \um\ line is 
$\sim 1.2\times 10^{-11}$ erg cm$^{-2}$ s$^{-1}$. The ratio \neratio$\sim
0.14$,
although this is uncertain since the lines appear in different, almost orthogonal, slits.
The spatial extent (FWHM) of the [Ne II] line-emitting region along the 
north-south-oriented SL slit is $5''$, while that of the 
dust-continuum-emitting region along the east-west oriented LL
slit is $12''$--$13''$. 
Considering the uncertainties in the size measurement with the slit spectrograph,
e.g., slight misalignment of the slit from the peak position, 
we consider that these results are consistent with the 
size of IRS1 determined from the \akari\ IRC image in \S~2.

We model the IRS spectrum as thermal emission 
from several independent dust components. 
We do not carry out a detailed analysis considering 
all possible dust species, which is 
beyond the scope of this paper. We instead 
determine the minimum number of dust species required to explain 
the observed IRS spectra and derive their characteristic parameters. 
The IRS spectrum in Figure 4 requires at least three dust components.
First, the narrow intensity peaks at 22.6, 27.2, and 33.5 \um\ clearly 
indicate a considerable amount of Mg-rich crystalline olivine, i.e.,
Mg$_{2x}$Fe$_{2-2x}$SiO$_4$ with $x\approx 1$.
The peak frequency and the relative strength of 
the mass absorption coefficient (hereafter MAC) 
depend on the Fe/Mg ratio, particle shape, particle size, and temperature 
\citep{fabian01, koike03, koike06}. The strong peaks at the above wavelengths 
are characteristics of Mg-rich, Fe-poor olivines 
\citep[see Fig. 1 of][]{koike03}.   
We adopt Mg$_{1.9}$Fe$_{0.1}$SiO$_4$ and use the optical constants
of \cite{fabian01}. 
We assume spherical dust grains 
with uniform radii of 0.1 \um\ and calculate their MAC following \cite{bohren83}. 
Second, the broad bump steeply rising from 15 \um\ and flattening 
around 18 \um\ could be attributed to   
metal oxides (Fe$_x$Mg$_{1-x}$O, $0 \le x \le 1$). 
Metal oxides have single-peaked, broad MACs
with peak wavelengths between 16 and 22 \um\ depending on 
the Fe content \citep{henning95}.
We use a combination of FeO and Mg$_{0.6}$Fe$_{0.4}$O
with optical constants of \cite{henning95}.
The MACs of 0.1 \um\ spherical particles, however, are
rather narrow to fit the observed broad bump, and 
we use the MAC for continuous distribution of ellipsoids for FeO
using a quadratic weighting where near-spherical particle shapes
are most probable.  
Different combinations of metal oxides might fit the 
spectrum equally well. For example, MgO has an absorption peak 
at wavelengths shorter than that of FeO and can be used instead 
of Mg$_{0.6}$Fe$_{0.4}$O (see below).
Finally, we need an additional, cool dust component 
for the far-IR emission that peaks 
at about 60 \um. Since we have only broad-band fluxes at $\ge 35$~\um, 
we cannot constrain the composition of this component. We adopt 
amorphous silicate, and use the optical constants
of cool silicate grains derived by \cite{suh99}.

With the above four dust components, 
i.e., crystalline olivine, FeO, Mg$_{0.6}$Fe$_{0.4}$O, and amorphous silicate, 
we fit the \spitzer\ IRS spectrum using MPFIT, which is a 
least-squares fitting tool based on the
Levenberg-Marquardt algorithm \citep{markwardt09}.
In the fit, we leave the temperatures and the masses of individual components 
free with reasonable lower and upper boundaries. 
In order to constrain the far-IR part of the spectrum, we use the \akari\ 65 and 90~\um\ fluxes. 
The best-fit masses and temperatures 
obtained by least-squares fitting are given in Table 2, where 
we also show the results when MgO is used instead of Mg$_{0.6}$Fe$_{0.4}$O (Model 2). 
The resulting model spectrum fits the observed IRS spectrum and also the 
far-IR broad-band fluxes reasonably well (Figs. 2 and 5).
The other model (Model 2) can fit the spectrum 
equally well (Fig. 5).
We could not exactly reproduce the observed spectrum because 
of the simplicity of the adopted dust model and because the dust components
are unlikely to be isothermal.
The total dust mass is $9\times 10^{-3}\dfour^2$~\msol, 
which is considerably greater than the estimate in \S~2 because of the 
lower temperature of the silicate dust components. 
Most of the mass comes from amorphous silicate, the cool dust component for the 
far-IR emission, while crystalline silicates contribute about 25\%.
The temperatures and therefore the masses 
of these two dust components vary little between the two models.
However, the parameters of the metal oxide components vary significantly, 
as can be seen from Table 2,
so they are not reliable. 
The temperature of crystalline silicate (55 K) 
is much lower than that (210 K) in HD 100546 \citep{malfait98}. The 
corresponding temperature in R71 is not available, but the temperature 
of the dust responsible for most of the emission at 10--100~\um\ in R71
has been estimated to be 120 K \citep{boyer10}. In R71, it was noted that 
an additional warm dust at 300 K is needed to explain the bump near 
10 \um\ \citep{voors99}.

\section{Molecular Emission Lines and Other Wavebands}

We carried out molecular line observations of 
\irasx\ using the 22-m telescope at the Mopra 
Observatory and \aste.
With Mopra, we mapped a
$6'\times6'$ area surrounding \irasx\ in the CO J=1--0 line
at 115.27 GHz (FWHM=$33''$) on June 28 and July 13, 2008. 
With ASTE, we made 5-point
cross observations at 20$''$ spacing in the CO J=3--2 line at 
345.79 (FWHM=22$''$) from 31 May to 6 June 2008.
We detected molecular gas at two LSR velocities, $-62$~\kms\ and +22 \kms\ surrounding 
\irasx\ 
(Figs. 6 and 7). Neither component peaks at the position of IRS1. 
Instead, the $-62$~\kms-component shows a cavity at the position of IRS1
while the +22 \kms-component shows no obvious spatial correlation.
The two clouds have a CO $(J=3-2)/(J=1-0)$ ratio $\sim 0.5$,
which is typical for giant molecular clouds.

We have also searched the 1.4 GHz continuum data of this area 
obtained from the Australian Compact Telescope Array
(Gaensler et al. 1999) and a deep (250 ks) Chandra X-ray archival observation,
but could not identify a counterpart. 
The upper limit ($5\sigma$) on the 1.4 GHz continuum flux is $\sim 10$~mJy.

\section{Nature of \irasx}

The strong crystalline silicate emission 
indicates that 
the dust in IRS1 is circumstellar or perhaps of supernova origin.
The steeply-rising MIR spectrum, 
the low-velocity [Ne II] 12.8 \um\ line, 
and the large extinction through the Galactic plane in
this direction almost rule out 
an extragalactic origin for \irasx.
The derived dust mass (Table 2) and the spatial 
extent ($0.19\times 0.10 d_4$ pc) imply a small self-extinction ($\simlt 0.1$ mag at K band);
together with the absence of an embedded star in images of \irasx, this rules out the possibility that it is an extended dust clump
powered by an embedded star.
It is unlikely that several highly-obscured stellar sources form IRS1, 
since the temperature (55--57 K) of the major dust components 
is less than expected for such circumstellar dust.
Instead, the [Ne II] and [Ne III] emission lines and the IR luminosity suggest that
it could be powered by a nearby hot, luminous star.
According to stellar atmosphere models, 
the ratio \neratio\ $\sim 0.14$ implies a stellar temperature of
30,000--40,000~K
\citep{morisset04b}.
The observed [Ne II] luminosity
implies a Lyman continuum photon luminosity $S_{\rm IRS1}= 
1.2\times 10^{47}\dfour^2$ s$^{-1}$ incident upon IRS1
\citep{ho07}.

The nearby O star Muzzio 10 can provide both the required luminosity and the
ionizing photons.
Based on new optical spectra,
\cite{bessell11} concluded
that its spectral type is O4.5III(fp)
and it is probably a member of the Cir OB1 association at $\sim 4$ kpc.
We adopt $T_*=40,500$ K, $L_*=5.8\times 10^5$~\lsol, and 
$S_*=3.2\times10^{49}$ s$^{-1}$ for Muzzio 10,
the parameters of an O4.5III star \citep{martins05}.
Note that the stellar temperature is consistent with that needed to
provide the observed [Ne III]/[Ne II] line ratio.
The projected distance of IRS1 from Muzzio 10 is 
13$''.7$, or $0.27\dfour$ pc, 
comparable to the size of IRS1.
At 4 kpc,  $L_{\rm IRS1}/L_*=3.4\times 10^{-3}\dfour^2$ and
$S_{\rm IRS1}/S_*=3.8\times 10^{-3}\dfour^2$, which are in excellent agreement.
If the projected surface area of IRS1 as seen from Muzzio 10 is the same
as that seen from Earth, then they are separated by 0.6 pc.
The expected temperature of amorphous silicate grains of radius
0.1 \micron\ is 70 K \citep{tielens05}, but 
the {\em mean} temperature of dust in a cloud with a large UV optical depth 
will be lower. 
The temperature of 57 K inferred in \S~3 (Table 2) therefore is consistent with 
our expectation.
For metal oxide dust, however, the expected temperature 
is considerably lower than the derived temperature (150 K) in \S~3.
This, however, could be due to our assumed dust composition, e.g., MgO instead of 
Mg$_{0.6}$Fe$_{0.4}$O yields 90--120 K instead of 150 K (Table 2). 
We conclude that the agreement between the predicted and observed
Ne line luminosities, IR luminosities and silicate dust temperatures supports the hypothesis
that IRS1 is powered by Muzzio 10,
although the high temperature of the small amount of the 
metal oxide dust remains unexplained. 
This association implies that IRS1 cannot be too old, since the ionizing radiation
from Muzzio 10 would have destroyed it 
in $\sim 10^5$~yrs \citep{bertoldi90}; it would also be subject to
destruction by the stellar wind from Muzzio 10.

What is the origin of the crystalline silicate? It can be produced
in both YSOs and evolved stars, since it requires temperatures $\ga 1000$~K to
form by thermal annealing
\citep{hen10}. 
The dust in YSOs is confined to circumstellar disks and
is thus very compact; hence, a YSO origin does not apply to
the present source, which is extended.
The absence of dense molecular material associated with \irasx\ is consistent with
this conclusion.
There are no luminous evolved stars within the boundaries of IRS1:
Muzzio 10 is very hot and must still be close to the main sequence; it is not
in the blue-loop phase of late-stage evolution of massive stars, and
its location in the H-R diagram is far from the 
``S Dor instability strip'' that defines the left boundary of the LBVs \citep{vink09}. 
There is a Wolf Rayet star (WR65; Oskinova \& Hamann 2008) and another 
possible mass-losing star in the field, but 
they are relatively distant
($4.'3$ south of and $4'.0$ north of IRS1 respectively) and 
show very different MIR colors (see Fig. 1). 
The only candidate for the source is the SNR \msh, which is at 
a comparable distance ($5.2\pm1.4$ kpc---\citealp{gaensler99}), 
{\em or its progenitor.} 
If we fit the broadband SED of Muzzio 10 from optical to MIR 
assuming a spectral type of O4.5III, 
we obtain a distance of 4.5 kpc and an absorbing H-nucleus column density
of $9.2\times 10^{21}$~cm$^{-2}$. This column density matches well with
that to the SNR determined from X-rays, $9.5\pm 0.3\times 10^{21}$~cm$^{-2}$ 
\citep{gaensler02}, which supports the association.
The crystalline silicates could have been produced either
in the late stages of the progenitor's evolution or in the cooling SN ejecta.
We favor a progenitor origin for IRS1 in view of its low velocity, 
$-160\pm 560$~km~s$^{-1}$. 
An origin in the progenitor is consistent with the ubiquity of crystalline silicates 
in evolved stars. Crystalline silicate emission features have not been detected
in the spectra of freshly-formed SN dust, although 
crystalline silicates of core-collapse SN origin have been discovered in
interplanetary dust particles \citep{messenger05}.
Crystalline silicates are subject to destruction and 
amorphization by SNR shocks and cosmic rays \citep[][and references therein]{jager03}, 
so that they will not survive long in the hostile environment of a young SNR
even if they are formed in the SN ejecta. 

A progenitor origin for IRS1 has several interesting consequences.
First, IRS1 should be embedded in the SNR and be relatively close 
to the explosion center, because otherwise, considering the 
large radius \citep[23 pc at 4 kpc;][]{gaensler99} of the remnant, 
the dust mass lost by the progenitor would be unacceptably large. 
This raises the question of how IRS1 could have survived the SN blast wave,
because most dust grains are destroyed by shocks faster than $\sim 200$~\kms\
\citep{jones96}.
One possibility would be that the explosion was very asymmetric so that 
IRS1 avoided the expanding dense ejecta. 
Another, more intriguing 
possibility is 
that the progenitor and Muzzio 10 formed a close binary system,
as first proposed by \cite{gaensler99}, and that Muzzio 10 
shielded IRS1 from the SN blast wave as shown, for example, 
in numerical simulations of Type Ia SN explosions in binary systems
\citep{marietta00}. 
The close proximity of Muzzio 10 and the pulsar B1509$-$58 
gives support to this possibility.
Since Muzzio 10 is currently an O4.5III star of 45 \msol \citep{martins05}, 
the progenitor should have had a very large (60--70 \msol) initial mass 
and a separation $\sim 100 R_\odot$ from Muzzio 10
(S.-C. Yoon, private communication). 
The progenitor could have lost most of its mass during Wolf-Rayet phase
prior to its explosion as an SN.
The shielding scenario for IRS1 implies a distance from the SN: If we assume 
that the diameter of Muzzio 10 in a binary was comparable to that of an 
isolated O4.5III star, i.e., $30 R_\odot$ \citep{martins05}, then it 
would shield a region 0.2 pc in extent (the observed size of IRS1) 
from a supernova $100 R_\odot$ away if IRS1 were about 0.6 pc from
the SN.
This is the same as the estimate of the current separation between
Muzzio 10 and IRS1 made above. 
Dust at this distance from the supernova could readily survive
the SN, as can be inferred from \citet{draine10}.
If a significant fraction of the mass loss that led to the formation of IRS1 
were in a disk of that radius, then the total dust mass of the disk would be 
about $(2\pi \times 0.6$~pc)/(0.2~pc) times the mass of IRS1, or 
about $0.17 M_\odot.$ 
For a gas to dust ratio of 100, this implies a total 
mass of $17 M_\odot$, which is large but not unreasonable.

Measurement of the proper motions of the pulsar and Muzzio 10 
would be the key test of 
this binary shielding model.
The orbits of the neutron star and its companion 
after the supernova explosion depend on 
their separation, masses, and the magnitude of the kick velocity 
imparted to the neutron star during the explosion \citep{tauris98}.
Since Muzzio 10 is much more massive than the neutron star, its velocity would
be almost unchanged by the SN, and as a result its motion should have been nearly perpendicular to the vector from the SN to IRS1.
The motion of the pulsar, on the other hand, could be mainly determined by the 
magnitude of the kick velocity and so could be in any direction. 
The available radio data on the pulsar B1509$-$58, however, 
do not show an appreciable proper motion 
(M. Livingstone, personal communication).
Determination that the progenitor was a member of
a binary system, together with the suggestion
by \cite{gaensler99} that the supernova was of Type Ib/c, 
would also support the binary model for SN Ib/c \citep[e.g.,][]{yoon10}, 
although this would require revision of the canonical view that only fast 
winds are expected in the final evolutionary stages of Type Ib/c progenitors 
\citep{che05}.
Above, we commented that the mid-IR spectrum of \irasx\ was similar to that of
an LBV, which experiences
extreme mass loss before its explosion and  
is associated with SN IIn supernovae \citep{smith10}. However, \cite{bessell11} has
obtained optical spectra toward IRS1 and detected no H$\alpha$ emission.
The absence of emission from hydrogen in the mass lost from the progenitor
supports a Type Ib/c origin rather than a Type IIn origin for the supernova.

The nature of the diffuse, extended structures surrounding IRS1 is uncertain. 
They are not correlated with the X-ray knots or arcs of the 
pulsar wind nebula \citep{gaensler02}.
\spitzer\ IRS spectra of these structures
indicate that they are a mixture of dust-continuum dominated 
material like IRS1 and line-dominated, ionized gas. It is likely that the red/yellow knots, spurs, 
and filaments in Figure 1 are dominated by dust continuum, while the diffuse green structures are 
dominated by line emission, in which case the former might be the progenitor's circumstellar material 
while the latter are SN ejecta. 
In one faint, [Ne II]-emitting gaseous clump, we measured a
central velocity $\sim +1,000$ \kms, consistent with SN ejecta.
The SL and LL slits, which are almost orthogonal to each other, 
provide only limited spectral and spatial information to reach further conclusions.
The spatial relation of IRS1 and the surrounding structures 
to the SNR as well as to Muzzio 10 needs further study.

\irasx\ appears to be the first case in which crystalline silicates have been 
observed to be associated with a SNR or its progenitor.
IRS1 and the associated structures may be the relics of massive
star death in a close binary system, 
as shaped by the supernova, the pulsar wind and the intense ionizing radiation
of the embedded O star. If confirmed by observations of the proper motion of
Muzzio 10 and pulsar B1509-58, we have a unique opportunity to study the
interior of a young SNR containing crystalline dust 
and illuminated by a central ionizing source.

\acknowledgements

This work is in part based on observations with AKARI, a JAXA project with
ESA,
and with the Spitzer Space Telescope,
operated by the Jet Propulsion Laboratory, California Institute of
Technology under a contract with NASA.
Support for this work was partially
provided by NASA through an award issued by JPL/Caltech.
We thank Ko Arimatsu for checking the possible effect of optical ghosts 
in the L15 image. 
We thank Matthew Bailes, Alex Filippenko, Vicky Kaspi, Maggie Livingstone,
Nathan Smith, and Sung-Chul Yoon for their helpful discussions.
We also thank the anonymous referee for his/her comments which considerably improved the presentation of the paper.
BCK was supported by the National Research Foundation of Korea
(NRF) grant funded by the Korea Government
(MEST---No. R01-2007-000-20336-0, NRF-2010-616-C00020). 
CFM was supported by NSF grant AST-0908553 and by the 
Groupement d'Int\'er\^et Scientifique
(GIS) ``Physique des deux infinis (P2I)."
The Mopra Telescope and ATCA are both parts of the Australia Telescope
supported by the Commonwealth of Australia
as a National Facility managed by CSIRO. The University of New South Wales
Mopra Spectrometer Digital Filter Bank
was provided with support from the Australian Research
Council, together with the University of New South Wales,
University of Sydney and Monash University.
The ASTE project is operated by Nobeyama Radio Observatory (NRO), a branch of National
Astronomical Observatory of Japan (NAOJ), in collaboration with University of Chile,
University of Tokyo, Nagoya University, Osaka Prefecture
University, Ibaraki University, and Hokkaido University.
Observations with ASTE were in part carried out remotely from Japan by using
NTT's GEMnet2 and its partner
Research and Education networks, which are based on
AccessNova collaboration of University of Chile, NTT Laboratories, and NAOJ.

{}

\begin{figure}
\epsscale{1.0}
\plotone{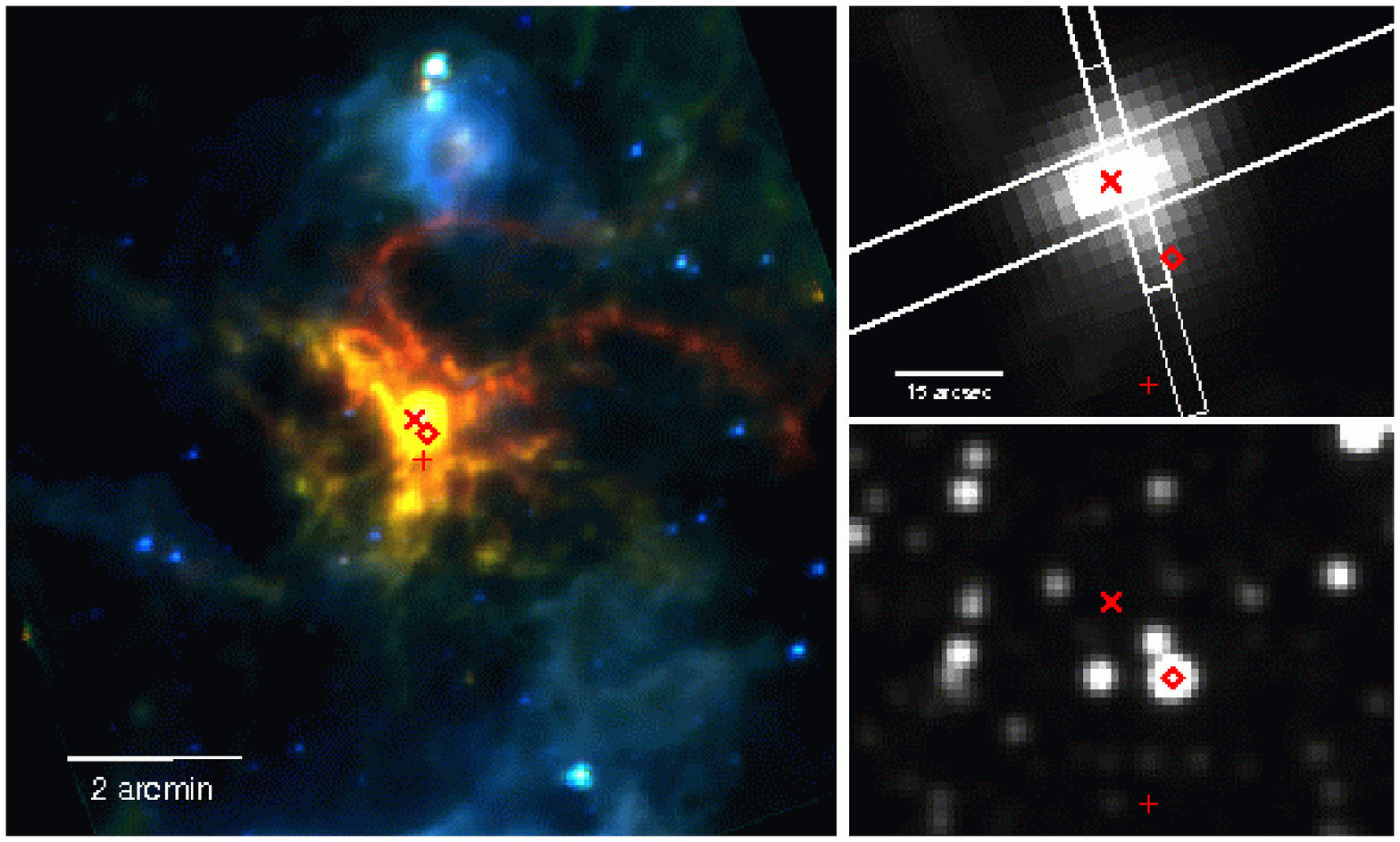}
\caption{
(left) Three-color image of \irasx\ produced from 
\akari\ S11 (blue), L15 (green), and L24 (red) images.
North is up and east is to the left. 
The brightness is adjusted to show the faint emission features, which
saturated the central part. 
The cross marks the peak position of the central compact source (IRS1)
at 15~\um, $(\alphaJ,\deltaJ)=(15^{\rm h} 13^{\rm m} 56^{\rm s}.32,
-59^\circ 07' 40.''9)$.
The plus and diamond signs mark the positions of the 
pulsar B1509$-$58 \citep{gaensler02}
and the O star Muzzio 10 ({\em 2MASS} 15135520$-$5907516), respectively. 
The bright star embedded in the diffuse blue emission 
in the south is the Wolf Rayet star WR65
while the other bright star of similar morphology in the north
is not previously known. 
(right)   
Magnified view of the central area in L15 band (upper panel) 
and the corresponding \tmass\ image (lower panel). 
The Spitzer IRS slit positions are shown by rectangular boxes
in the upper panel; 
large ($148''\times 10.7''$) solid one representing LL band slit
position and two superposed small ($50.4''\times 3.7''$)
solid ones representing the two nodded SL band slit positions.
}
\end{figure} 

\begin{figure}
\epsscale{1.0}
\plotone{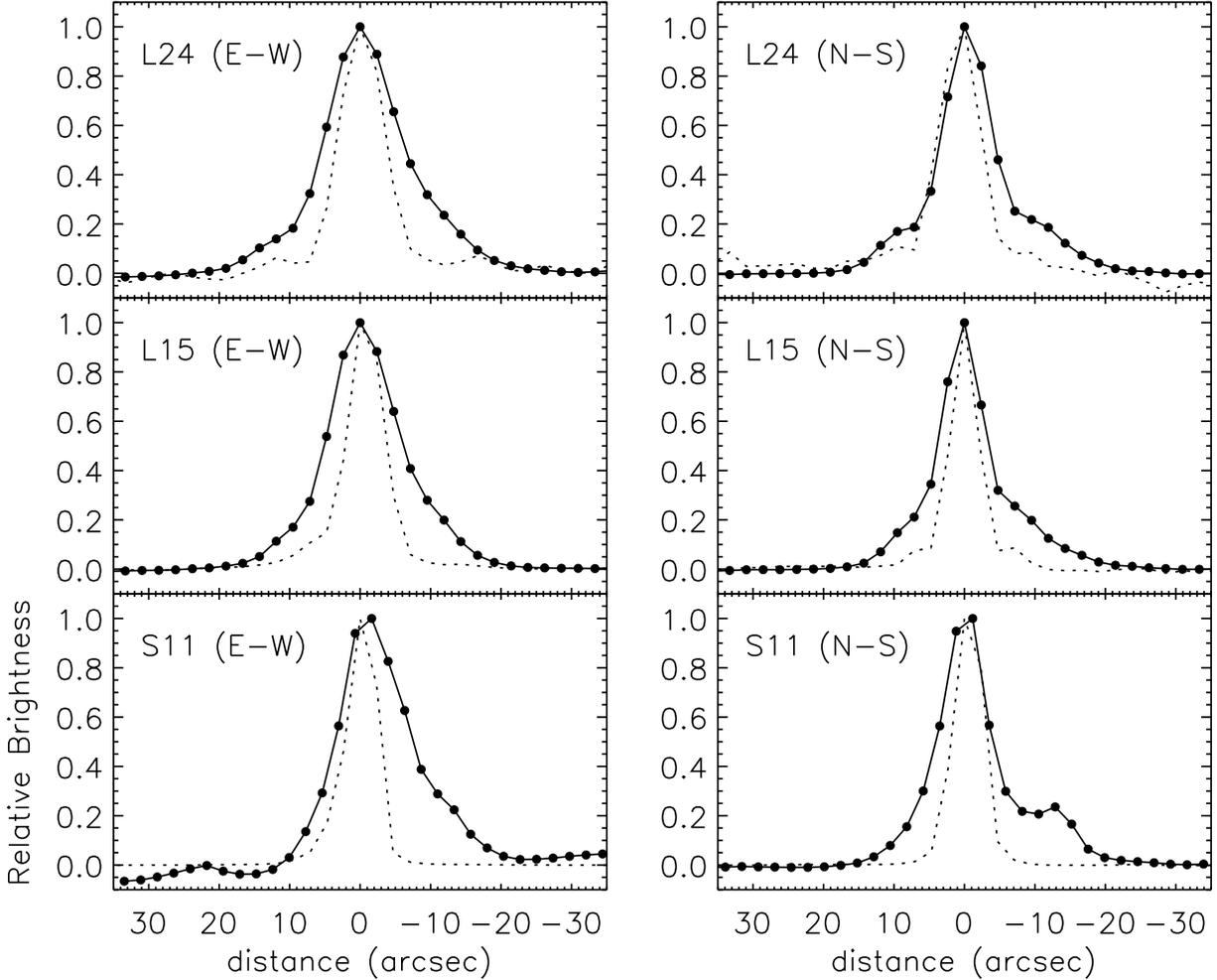}
\caption{
One-dimensional, normalized intensity profiles of the central compact source IRS1
at 24, 15, and 11 \um\ (solid lines). The left panels show the profiles
along the east-west direction (P.A.=110$^\circ$) while the right panels
show the profiles perpendicular to it.
The $(0, 0)$ position represents the peak position at
15~\um, and the filled circles represent the intensities at individual pixels.
The 24~\um\ image has a relatively large uncertainty in astrometry, and therefore
the profile is shifted to match the peak position at 15~\um.
The pixel sizes are $2.''34$--$2.''38$.
The dotted lines show the profiles of a point source in the same images.
}
\end{figure}

\begin{figure}
\epsscale{0.6}
\plotone{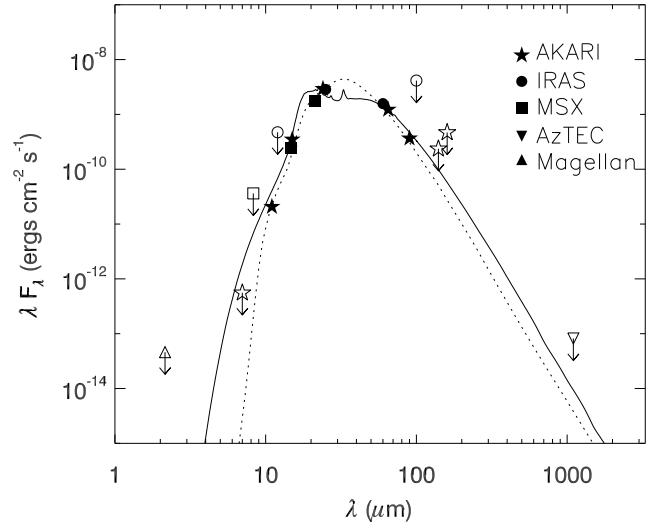}
\caption{
Spectral energy distribution of IRS1 of
\irasx. The open symbols represent upper limits.
The fluxes are color corrected (see Table 1).
The dotted line is a model SED of thermal dust emission at 79 K, whereas
the solid line is a model SED resulting from a fit to the \spitzer\ IRS spectrum 
(Model 1 in Table 2; see \S~3).
}
\end{figure}

\begin{figure}
\epsscale{0.8}
\plotone{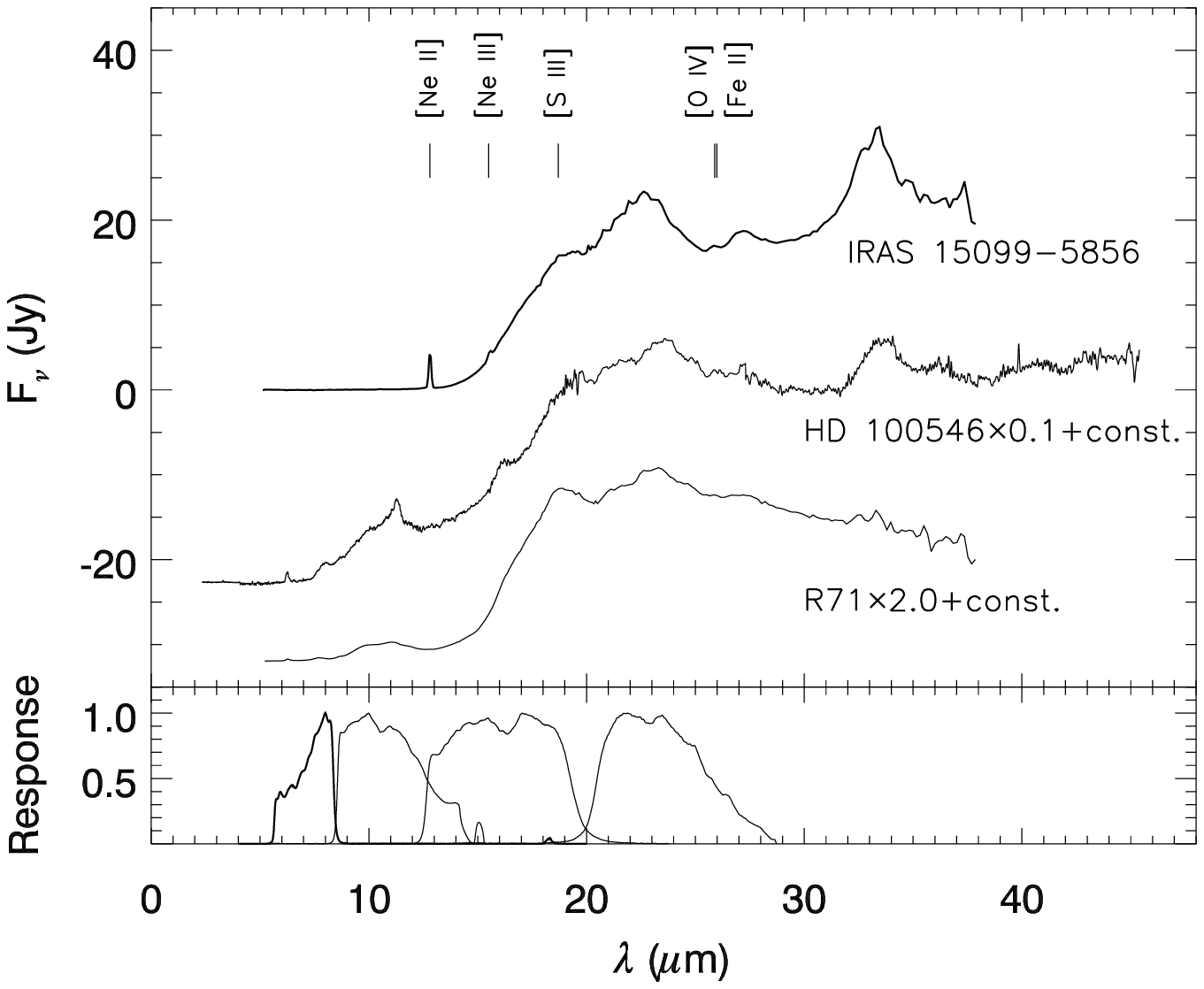}
\caption{
Spitzer IRS spectrum of \irasx. 
Wavelengths of the identified ionic lines are marked. 
The $\ge 15$ \um-part of the \irasx\ spectrum is very similar to the spectrum of 
the Herbig Be (B9.5e) star HD 100546 \citep{malfait98} and 
the LBV R71 (HD 269006) in the LMC \citep{voors99, morris08},
which are shown for comparison. The HD 100546 spectrum is from
the ISO Short Wavelength Spectrometer spectral atlas of \citet{sloan03b} 
while the R71 psectrum is from the Spitzer Heritage Archive.
The bottom panel shows the relative responses of the \akari\ 
S7, S11, L15, and L24 imaging bands.
}
\end{figure}

\begin{figure}
\epsscale{1.0}
\plotone{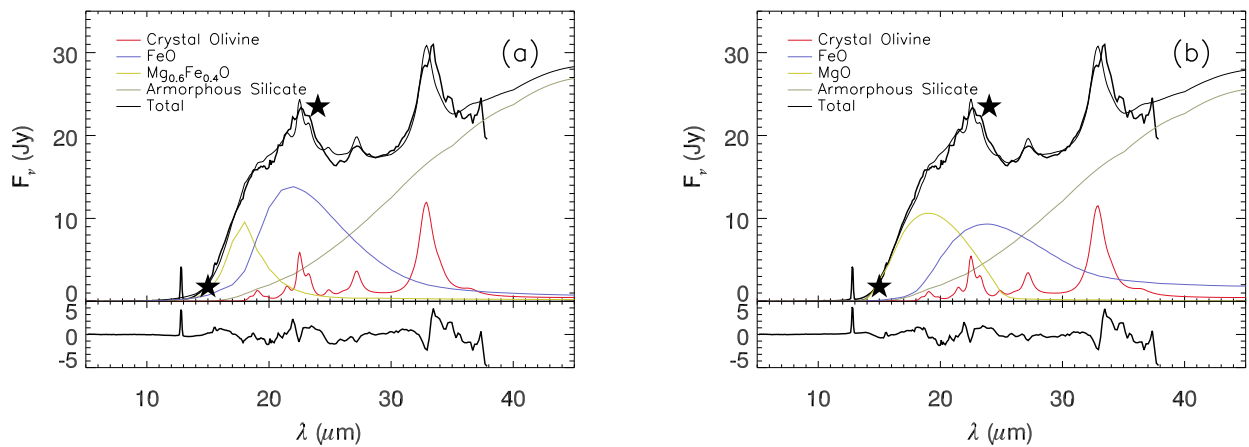}
\caption{
Model fits to the Spitzer IRS spectrum of \irasx\ (thick solid lines); (a) Model 1 and (b) Model 2 in Table 2, respectively. 
The thin solid line in each plot represents a model of 
thermal emission from several independent dust components,
which are shown. 
The filled-star symbols mark the 15 and 24 \um\ \akari\ fluxes.
The bottom frames show the differences between the model fits and the observed spectrum. 
}
\end{figure}

\begin{figure}
\epsscale{1.0}
\plotone{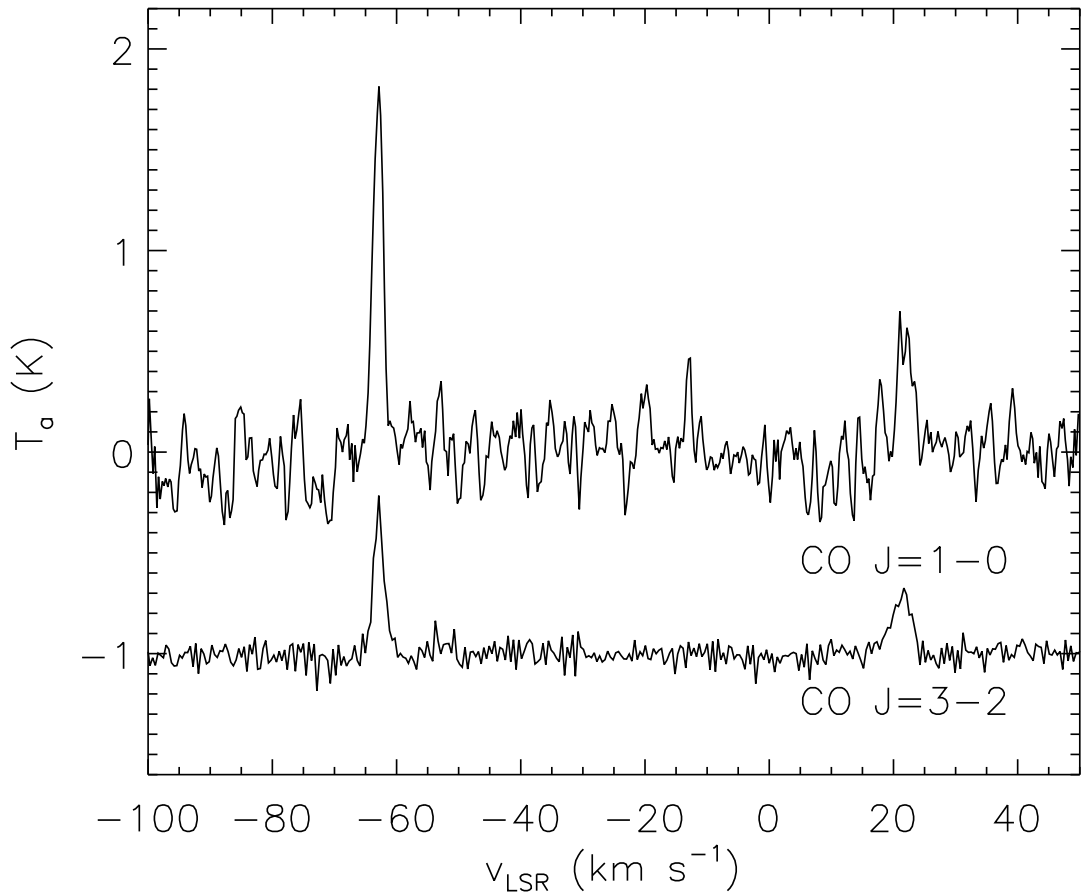}
\caption{
CO $J=1-0$ and $J=3-2$ line profiles toward IRS1. The 
former is obtained by the 22-m Mopra telescope (FWHM=$33''$) 
while the latter is obtained by 
the 10-m ASTE telescope (FWHM=$22''$).
}
\end{figure}
\clearpage

\begin{figure}
\epsscale{1.0}
\plotone{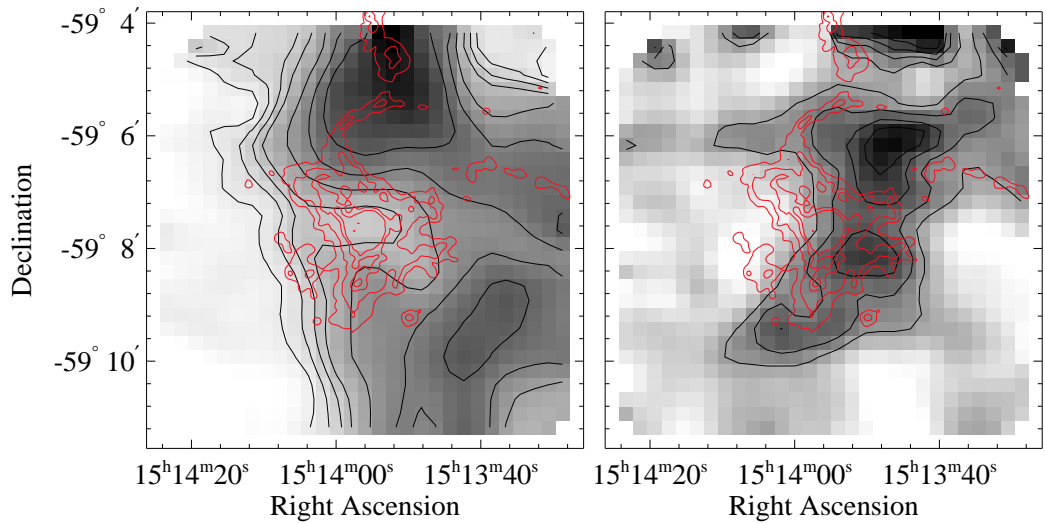}
\caption{
CO $J=1-0$ line intensity distribution of two molecular clouds
detected toward \irasx; the $-62$~\kms\ ($-65$ to $-61$~\kms) 
and +22 \kms\ (+16 to +25 \kms) clouds are in the left and
right frames, respectively. 
Contour levels representing integrated intensity are
20, 30, 40, 50, 70, 90, and 110 K \kms.
The gray-scale plot distinguishes hills from valleys. 
The red contours show the 24 \um\ brightness of \irasx.}
\end{figure}
\clearpage

\clearpage

\begin{deluxetable}{lccc}
\tablecaption{Flux of IRAS 15099$-$5856 IRS1 \label{tbl-1}}
\tablewidth{0pc}
\tablecolumns{4}
\tablehead{
\colhead {} & \colhead{} & \multicolumn{2}{c}{Flux} \\
\cline{3-4} \\
& \colhead {Wavelength} & \colhead{Observed} & {Color-Corrected} \\
\colhead {Telescope} & \colhead {($\mu$m)} & \colhead{(Jy)} & \colhead{(Jy)}
}

\startdata

Magellan 	&  2.15  & $\le 7\times 10^{-6}$ & ... \\
\\
MSX      	& 8.28   & $\le 0.101^a $  &  ...\\
                & 14.65  &  1.43 (0.10)  & 1.20 (0.08) \\
             	& 21.3   & 12.88 (0.77)  & 12.50 (0.75) \\
\\
AKARI IRC       &  7     & $\le 0.0013$ & ... \\
             	& 11     & 0.116 (0.003)  & 0.076 (0.002)\\
             	& 15     & 5.14 (0.15)    &  1.75 (0.02) \\
             	& 24     & 21.4 (1.0)     & 23.5 (1.1) \\
\\
AKARI FIS       & 65     & 27.9 (4.8) & 26.6 (4.6) \\
             	& 90     & 17.2 (2.5) & 11.0 (1.6) \\
             	& 140    & $\le 11^b$ & ... \\
             	& 160    & $\le 25^b$ & ... \\
\\
IRAS     	& 12     & $\le 1.89$ & ...\\
             	& 25     & 19.9 (2.0)   & 23.8 (2.4) \\
             	& 60     & 37.0 (4.1)   & 31.5 (3.5) \\
             	& 100    & $\le 138^c$  & ... \\
\\
AzTEC    	&1100    & $\le 0.03$ & ... \\

\enddata

\tablecomments{Magellan, AKARI IRC, and AzTEC fluxes are from this work. 
For the non-detections, the listed fluxes are $2\sigma$ upper limits.
The other fluxes are from the Point Source Catalogs.}
\tablenotetext{a} {MSX flux limit with quality\_flag=1 \citep{egan03}.}
\tablenotetext{b} {Detection limits according to the FIS Bright Source Catalog 
Version 1.}
\tablenotetext{c} {Upper limits from the IRAS Point Source Catalog Version 2.0, which are nominally $3\sigma$ values.}

\end{deluxetable}
\clearpage

\begin{deluxetable}{ccccc}
\tablecaption{Dust Parameters of IRS1 from the \spitzer\ IRS spectrum \label{tbl-2}}
\tablewidth{0pc}
\tablecolumns{5}
\tablehead{
\colhead {} &  \multicolumn{2}{c}{Model 1} & \multicolumn{2}{c}{Model 2} \\
\cline{2-3} \cline{4-5} \\
& \colhead{Mass} & \colhead {Temperature} & \colhead{Mass} & {Temperature} \\
\colhead {Component} & \colhead{($10^{-3} \dfour^2$ \msol)} & 
 \colhead {(K)} & \colhead{($10^{-3} \dfour^2$ \msol)} & \colhead{(K)} 
}

\startdata

crystalline olivine 
(Mg$_{1.9}$Fe$_{0.1}$SiO$_4$) & 2.3 (0.7)       & 54.8 (1.6)    & 2.5 (0.7)     & 54.1 (1.6) \\
FeO                         & 0.020 (0.003)   & 146 (4)       & 0.22 (0.08)   & 89.4 (4.9) \\
MgO                         & ...             & ...           & 0.025 (0.006) & 116 (3) \\
Mg$_{0.6}$Fe$_{0.4}$O       & 0.0069 (0.0016) & 147 (6)       & ...           & ... \\
amorphous silicate          & 6.6 (0.3)       & 57.2 (0.4)    & 6.1 (0.2)     & 57.6 (0.4) \\
\cline{1-5} \\
Total                       & 8.9 (1.0)$^a$       & ...           & 8.9 (1.0)$^a$      & ... \\
\enddata

\tablenotetext{a} {
Since the estimates of the  mass of the different components are not independent, 
we give an upper limit on the error in the total mass by simply adding the 
errors of the individual components.}
\end{deluxetable}
\clearpage

\end{document}